\documentclass[twocol]{ametsoc}
\journal{jas}

\bibpunct{(}{)}{;}{a}{}{,}

\usepackage{dirtytalk, float}
\usepackage{amsmath}
\usepackage{xcolor}
\usepackage{mathtools}

\newcommand\Rey{\mbox{\textit{Re}}} 

\newcommand{\pd}[2]{\frac{\partial #1}{\partial #2}}
\renewcommand{\vec}[1]{\boldsymbol{#1}  } 
\renewcommand{\bar}{\overline}

\title{Buoyant Motion of a Turbulent Thermal}

\authors{Nathaniel Tarshish\correspondingauthor{Atmospheric and Oceanic Sciences Program, 300 Forrestal Road, Sayre Hall, Princeton, NJ USA}}

\affiliation{Atmospheric and Oceanic Sciences Program, Princeton University, Princeton, NJ 08544, USA}

\email{nathaniel.tarshish@noaa.gov}

\extraauthor{Nadir Jeevanjee}
\extraaffil{Department of Geosciences, Princeton University, Princeton, NJ, 08544, USA}

\extraauthor{Daniel Lecoanet}
\extraaffil{Princeton Center for Theoretical Science, Princeton University, Princeton, NJ 08544, USA}

\abstract{By introducing an equivalence between magnetostatics and the equations governing buoyant motion, we derive analytical expressions for the  acceleration of isolated density anomalies, a.k.a. thermals. In particular, we investigate buoyant acceleration, defined as the sum of the Archimedean buoyancy $B$ and an associated perturbation pressure gradient. For the case of a uniform spherical thermal, the anomaly fluid accelerates at $2B/3$,  extending the textbook result for the induced mass of a solid sphere to the case of a fluid sphere. For a more general ellipsoidal thermal, we show that the buoyant acceleration is a simple analytical function of the ellipsoid's aspect ratio. The relevance of these idealized uniform-density results to turbulent thermals is explored by analyzing direct numerical simulations of thermals at $\Rey =  6300$. We find that our results fully characterize a thermal's initial motion over a distance comparable to its length. Beyond this buoyancy-dominated regime, a thermal develops an ellipsoidal vortex circulation and begins to entrain environmental fluid. Our analytical expressions do not describe the total acceleration of this mature thermal, but still accurately relate the buoyant acceleration to the thermal's mean Archimedean buoyancy and aspect ratio. Thus, our analytical formulae provide a simple and direct means of estimating the buoyant acceleration of turbulent thermals.} 		

\begin{document}

\maketitle
\section{Introduction}
\label{sec:intro}
Buoyancy is the root cause of thermal convection in the atmosphere and fundamentally governs the vertical acceleration of clouds. A density anomaly's buoyancy is most often characterized by the Archimedean formula $B = - \rho'/\rho_og$. However, only an \emph{infinitesimal} buoyant parcel, whose motion has a negligible impact on the environment, actually accelerates at its familiar Archimedean value. A density anomaly of finite size disrupts its environment and sets the surrounding fluid in motion. This loss of momentum to the environment reduces the finite parcel's buoyant acceleration from its Archimedean buoyancy to its "effective buoyancy" $\beta$, as introduced by \cite{davies2003} and discussed by \cite{doswell_markowski_2004,xu_randall_2001,krueger_mclean_fu_1995}. The back-reaction exerted by the environment on the buoyant fluid is realized by a buoyancy perturbation pressure (to be fully defined in the next section).

The reduction from $B$ to $\beta$ is determined by the geometry of the density anomaly. As discussed in \cite{houze2014cloud},  for broad pancake-shaped density anomalies, buoyant motion is severely inhibited by the surrounding fluid, and $\beta$ can be several orders of magnitude less than $B$. On the other hand, narrow, pencil-shaped density anomalies impart negligible momentum to their environment, and $B$ serves as a reliable proxy for $\beta$. 

Moving beyond this qualitative understanding, approximate formulae and scalings exist for the effective buoyancy in updrafts \citep{morrison_2016_1,morrison_2016_2,peters_2016}, of an isolated bubble \citep{pauluis_garner_2006}, and in hydrostatic versus non-hydrostatic models \citep{weisman_skamarock_klemp_1997,jeevanjee2017b}. Furthermore, exact analytical results for the effective buoyancy along the central axis of a buoyant cylinder were obtained by \cite{Jeevanjee2016}. While this progress is significant, we still lack a full 3D analytical solution for the effective buoyancy of a density anomaly even for idealized geometries, and thus have no quantitative picture for how the \emph{environment} responds to buoyant accelerations.

Another issue is that convection schemes often account for the buoyancy perturbation pressure gradient by simply multiplying $B$ by a fixed virtual (a.k.a. induced) mass coefficient. This coefficient is sometimes set equal to 2/3, which is the value for a solid sphere in potential flow (e.g., see \cite{romps2015sticky, romps2010undiluted,de2012parameterization,batchelor2000introduction}). In a similar fashion, \cite{turner_1964} approximated the perturbation pressure acting on an ellipsoidal thermal by appealing to the virtual mass coefficient of a solid ellipsoid. Not only does fixing a coefficient neglect the geometry-dependent nature of the buoyancy perturbation pressure, but there also does not seem to be any rigorous proof that the solid-body value is appropriate for a \emph{fluid} density anomaly, which may experience an internal circulation and thus deform as it accelerates.

In this paper, we aim to make progress on these questions by deriving the exact effective buoyancy for the idealized case of a uniformly buoyant ellipsoid. Our analytical results spring in part from the recognition of a novel mathematical equivalence between the equations of magnetostatics and effective buoyancy. We  show that the magnetic field of a uniformly magnetized body of arbitrary shape is mathematically identical to the effective buoyancy of an identically-shaped fluid density anomaly. 

This equivalence allows us to leverage the long-standing classical literature on magnetization to find the effective buoyancy of an ellipsoidal thermal. In addition, the correspondence provides theoretical justification for appealing to the solid-body virtual mass coefficients in the case of ellipsoids, yet also proves that the connection between solid-body motion and fluid anomaly acceleration breaks down for all other geometries.

The relevance of the ellipsoidal geometry examined in this work is supported by the observed shapes of convecting thermals. The actively convecting cores of Cumulus in Large Eddy Simulations \citep{sherwood_2013, romps2015sticky} and atmospheric observations \citep{damiani_2006, damiani_2007} are often ellipsoidal in shape. Laboratory experiments \citep[e.g.][]{scorer_1957,woodward_1959} of rising thermals also demonstrate this same ellipsoidal form.  

Realistic turbulent thermals, of course, have non-uniform $B$ fields.  How relevant are theoretical formulae for $\beta$ that assume uniform $B$? To answer this we present simulations of a turbulent thermal, and show that its \emph{average} $\beta$ can indeed be explained by our analytical formulae, when applied to the turbulent thermal's average $B$ field.

\section{Effective Buoyancy Preliminaries} 
\label{sec:beta_prelim}
We begin with the Boussinesq momentum equation,
\begin{equation}
\frac{D \vec{v}}{D t}=  - \frac{ \nabla p'}{\rho_o } + B \, \vec{e}_z + \nu \nabla^2 \vec{v} \, , 
\label{eqn:boussinesq} 
\end{equation}
where  $B = - \rho'/\rho_og$ is the Archimedean buoyancy and $\nu$ is the molecular viscosity \citep{markowski2011mesoscale}. We decompose the pressure according to $p'(x,y,z,t) = p(x,y,z,t) - \bar{p(z)}$,  where $\bar{p(z)}$ is in hydrostatic balance with the constant reference density, $\rho_o$. Consider the divergence of the momentum equation, 
\begin{equation}
\rho_o  \frac{\partial }{\partial t} \left[ \nabla \cdot \vec{v} \right] = - \rho_o \nabla \cdot \left( \vec{v} \cdot \nabla \vec{v} \right) - \nabla^2 p' + \rho_o \pd{B}{z} + \rho_o \nu \nabla^2 \left[ \nabla \cdot \vec{v} \right] \,. 
\end{equation} 
Boussinesq mass continuity ($\nabla \cdot \vec{v} = 0$) dictates that the total convergence tendency is zero, yielding a Poisson's equation for $p'$,  
\begin{equation}
\nabla^2 p'  = -  \rho_o \nabla \cdot \left( \vec{v} \cdot \nabla \vec{v} \right) + \rho_o\pd{B}{z} \, . \label{p'eqn}
\end{equation}
From this perspective, the perturbation pressure arises to enforce mass continuity by opposing any convergence tendencies generated by advective motion or buoyancy.  

The distinct physical mechanisms generating the perturbation pressure motivates further decomposing $p'$ according to $ p' = p_b + p_d $, where $p_b$ is the buoyancy perturbation pressure driven statically by vertical gradients in the buoyancy field, and $p_d$ is the dynamic pressure generated by advective motion. Linearity of Poisson's equation then allows us to separate Eqn. \ref{p'eqn} into independent parts,
\begin{subequations}
\begin{gather}
\nabla^2 p_b  = 
\rho_o\pd{B}{z} \label{eqn:poisson_buoyancy} \\  
\nabla^2 p_d  = 
- \rho_o \nabla \cdot \left( \vec{v} \cdot \nabla \vec{v} \right)  \label{eqn:poisson} \, . 
\end{gather}
\end{subequations}
Likewise, the Lagrangian acceleration admits a  complimentary decomposition. As in \cite{Jeevanjee2015}, we define the effective buoyancy $\beta$ as the Lagrangian vertical acceleration that would result from instantaneously zeroing out the velocity field, 
\begin{equation}
\beta = \frac{D w}{D t} \bigg|_{ \vec{v} \rightarrow 0} \,
= -  \frac{1}{ \rho_o } \pd{p_b}{z} + B \, . 
\label{eqn:beta_momentum}
\end{equation}
By accounting for $p_b$, the effective buoyancy is thus the true buoyant acceleration experienced by a finite-sized density anomaly.

We also define the dynamic acceleration as the Lagrangian acceleration that would result from instantaneously zeroing out the density anomaly field, 
\begin{equation}
a_d = \frac{D w}{D t} \bigg|_{\rho' \rightarrow 0}  =  - \frac{1}{ \rho_o } \pd{ p_d}{z} + \nu \nabla^2 w \, . 
\label{eqn:inert_accel}
\end{equation}
Together, the effective buoyancy and the inertial acceleration combine to give the Lagrangian vertical acceleration, 
\begin{equation}
\frac{D w}{D t} = \beta + a_d. 
\end{equation} 
Note that these definitions fully specify the boundary conditions necessary to determine $p_b$ and $p_d$ as detailed in \cite{Jeevanjee2015}. 

Our focus in this article is on the effective buoyancy, but this is not to suggest that the dynamic acceleration is of little importance in atmospheric convection. In a mature thermal, the dynamic acceleration is comparable in magnitude to the effective buoyancy and critical for establishing and maintaining the thermal's internal circulation. For the initial motion of a nascent thermal, however, the effective buoyancy dominates over the dynamic acceleration. In Section \ref{sec:motion_from_rest}, we further detail the transition between these two regimes.

\begin{figure*}
 \centerline{\includegraphics[width=7in]{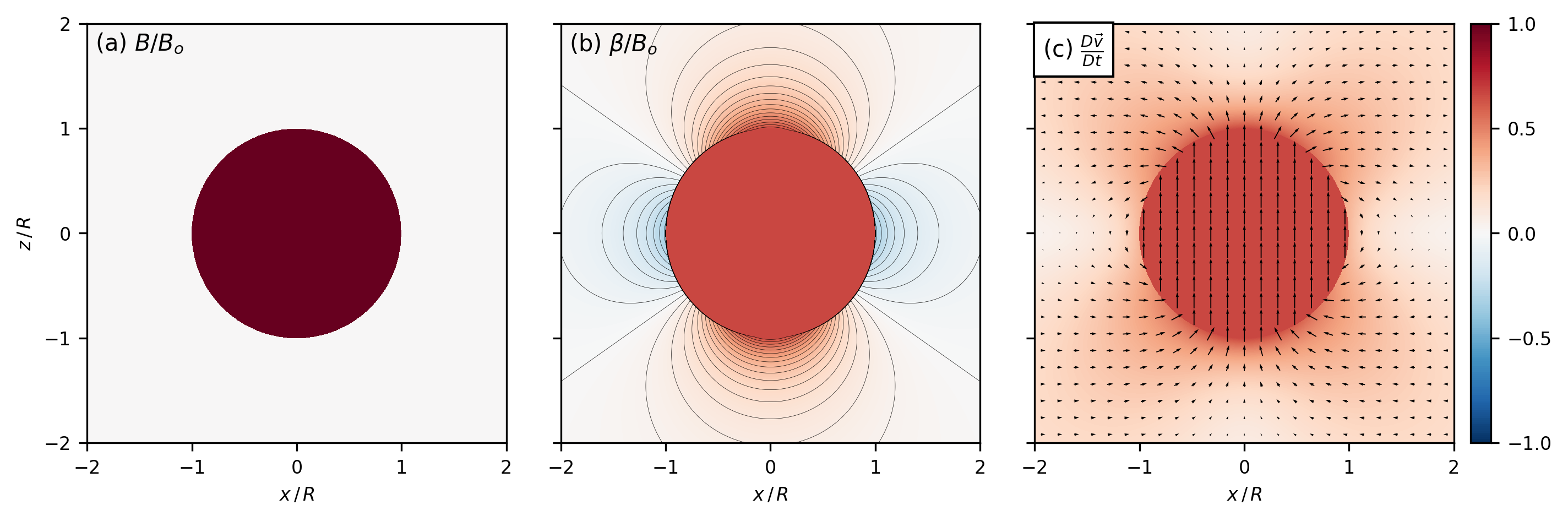}}
\caption{A uniform spherical thermal: a) vertical cross-sections of the Archimedean buoyancy profile, b) analytical solution for the effective buoyancy, and c) analytical solution for the Lagrangian acceleration with vector acceleration in arrows and acceleration magnitude in color.}\label{fig:sphere_plots}
\end{figure*}

\section{Effective Buoyancy of a Spherical Anomaly}
\label{sec:sphere_analytical_results}
We begin by seeking a solution for the effective buoyancy field of a spherical buoyancy anomaly of uniform Archimedean buoyancy $B_o$ and radius $R$. In spherical coordinates centered on the buoyant sphere with $\vec{r} = (r, \varphi, \theta)$, polar angle $\theta$, and azimuthal angle $\varphi$, the buoyancy field is described by 
\begin{equation}
B(\mathbf{r} ) = B_o \mbox{H} (R -r ) \, , 
\end{equation} 
where $\mbox{H}(R-r)$ is the Heaviside step function. Relying on Eqn. \ref{eqn:poisson_buoyancy},  we compute the buoyancy perturbation pressure as 
\begin{equation}
- \nabla^2 p_b  =    \rho_o B_o \cos{\theta} \, \delta (R -r )  \, \label{eqn:pb_sphere_poisson}
\end{equation}
where $\delta (R -r ) $ is the Dirac delta function. Away from $r=R$, $p_b$ solves Laplace's equation, and thus admits a series solution in the spherical harmonics. The conditions that $p_b$ be finite at $r=0$ and vanish at infinity provide the piece-wise expansion, 
\begin{equation}
p_b(r, \theta) = \left\{
\begin{array}{ll}
\displaystyle \sum_{\ell} a_{\ell} r^l P_{\ell}(\cos{\theta}) \,  & r<R \\[5pt]
\displaystyle  \sum_{\ell} \frac{ b_{\ell} } {r^{\ell+1} } P_{\ell}(\cos{\theta})  & r>R, 
\end{array} 
\right.
\end{equation}
where $P_{\ell}$ is the $\ell$-th member of the Legendre polynomials, which form a normalized basis for azimuthally-symmetric functions on the sphere. For future reference, we note that  $P_1(\cos \theta ) = \cos \theta$ and $P_2(\cos \theta) = 1/2 \left(3\cos^2 \theta - 1\right)$.  Dependence on $\varphi$ is precluded by azimuthal symmetry (whereas $\theta$ symmetry is broken by gravity).  
To generate the boundary conditions, we multiply Eqn. (\ref{eqn:pb_sphere_poisson}) by $r^2$ and separate $r^2\nabla^2$ into a radial and angular part,
\begin{equation} \pd{}{r}\left(r^2 \pd{p_b}{r} \right) 
+ \nabla^2_{\text{ang}} p_b =  - \rho_o B_o \cos{\theta} \,r^2 \delta (R -r ) \,, 
\label{eqn:boundary_cond_eqn}
\end{equation} 
where we have defined  
\begin{equation}
\nabla^2_{\text{ang}} \equiv
\frac{1}{\sin\theta} \frac{\partial}{\partial \theta}\left(\sin\theta \frac{\partial }{\partial \theta}\right) 
+ \frac{1}{\sin^2\theta} \frac{\partial^2 }{\partial \varphi^2} \, . 
\end{equation}
Suppose $p_b$ were discontinuous about $r=R$, then $\partial^2 p_b / \partial^2 r$ would generate a $\partial_r \left[ \delta (R - r) \right]$ on the L.H.S of Eqn. \ref{eqn:boundary_cond_eqn}. Since the singularities on the L.H.S and R.H.S. are now of different order (i.e., $\partial_r \left[ \delta (R - r) \right]$ is linearly independent from $\delta(R-r)$), we have a contradiction. Therefore, $p_b$ is continuous.

Integrating Eqn. \ref{eqn:boundary_cond_eqn} with respect to $r$ over $[R-\epsilon,R+\epsilon]$ then yields, 
\begin{equation}
\lim_{\epsilon \rightarrow 0 }  \, \left[ \,\pd{p_b}{ r } \bigg|_{R + \epsilon } - \, \pd{p_b}{ r }\bigg|_{R - \epsilon} \, \right]  =  - \rho_o B_o \cos{\theta} \,, 
\label{eqn:pb_bc}
\end{equation}
where the angular term dropped out as $\epsilon \rightarrow  0$ since $p_b$ is continuous in $r$ and smooth in $\theta$. 

Continuity of $p_b$ provides that $b_\ell = a_\ell R^{2\ell + 1}$, and we may write the expansion outside the spherical shell as
\begin{equation}
\sum_{\ell} \frac{ a_\ell R^{2 \ell + 1} } {r^{\ell+1} } P_{\ell}(\cos{\theta}) \,. 
\end{equation}
Enforcing the boundary condition (\ref{eqn:pb_bc}) on the radial derivative gives 
\begin{equation}
\lim_{\epsilon \rightarrow 0} \sum_{\ell=1}^{\infty}  \left[ \frac{ - (\ell + 1) R^{2\ell + 1} }{ (R + \epsilon)^{\ell+2 } }    - \ell  (R-\epsilon)^{\ell-1}  \right] a_\ell P_{\ell}(\cos{\theta})  = - \rho_o B_o \cos{\theta} . 
\end{equation}
Recognizing  $P_1(\cos{\theta})$ on the R.H.S and invoking the linear independence of the Legendre polynomials demands that any factors in front of the $l \neq 1$ Legendre polynomials are individually zero,
\begin{subequations}
	\begin{gather}
	\lim_{\epsilon \rightarrow 0} \left[ \frac{ - (\ell + 1) R^{2\ell + 1} }{ (R + \epsilon)^{\ell+2 } }    - \ell  (R-\epsilon)^{\ell-1}  \right] a_\ell   = 0 \\ 
	\implies \quad (2\ell + 1) R^{\ell - 1}\ a_\ell    = 0 \, .  
	\end{gather}
\end{subequations}
We find that  $a_\ell = 0  $ for $\ell \neq 1$. For the $\ell = 1$ case, we have 
\begin{subequations}
	\begin{gather}
	\lim_{\epsilon \rightarrow 0} \left[ \frac{ - 2 R^{3} }{ (R + \epsilon)^{3} }    - 1  \right] a_1   = -\rho_o B_o \\
	\implies \quad \ a_1    =  \frac{\rho_o B_o}{3},  
	\end{gather}
\end{subequations}
yielding the solution for the buoyancy perturbation pressure:
\begin{equation}
p_b(r, \theta) = \left\{
\begin{array}{ll}
\displaystyle \frac{ \rho_o B_o }{3} r \cos{\theta} \,  & r \leq R \\[9pt]
\displaystyle \frac{ \rho_o B_o R^3  }{3} \frac{\cos{\theta} }{r^2}  & r \geq R \,.  \label{eqn:pb_solution}
\end{array} 
\right.
\end{equation}
From Eqn. \ref{eqn:beta_momentum}, the effective buoyancy field readily follows:
\begin{equation}
\beta (\mathbf{r}) = \left\{
\begin{array}{ll}
\displaystyle \frac{2 B_o}{3} \,  & r < R \\[9pt]
\displaystyle  \frac{B_oR^3}{3} \frac{\left( 3 \cos^2 \theta  - 1 \right)}{r^3}  & r > R \,. 
\end{array}
\right.  \label{eqn:effective_buoyancy_sphere}
\end{equation}
This analytical expression for $\beta(\mathbf{r})$ is a main result of this paper, and is plotted in the center panel of Figure \ref{fig:sphere_plots}. Note the presence of $P_2(\cos{\theta})$ in Eqn. (\ref{eqn:effective_buoyancy_sphere}), with characteristic lobes evident in Fig. \ref{fig:sphere_plots}.   Furthermore, setting $\theta = \pi/2$ in Eqn. (\ref{eqn:effective_buoyancy_sphere}) yields a horizontal profile of $\beta$: the sphere itself accelerates vertically at $2/3B_o$, while the immediately adjacent environment subsides at $-\frac{1}{3}B$. This subsidence acceleration then decays fairly rapidly as $r^{-3}$.

One implication of these results is that the negative acceleration associated with near-field compensating subsidence is comparable in magnitude to the acceleration of the anomaly itself. This environmental acceleration declines rapidly with distance, yielding after finite time a descent of environmental air that is strongly enhanced near the anomaly. In a stably stratified environment (i.e., $d\theta/dz >0$), such a distribution of descent yields a corresponding distribution of positive buoyancy anomalies, which then generate their own convection and cause this chain to repeat itself. This is, of course, simply a description of how gravity waves propagate away from rising fluid, and we see that the horizontal inhomogeneity of the environmental acceleration evident in Fig. \ref{fig:sphere_plots} is critical to this propagation.

For completeness, we also note that the environmental fluid's horizontal acceleration can be computed as
\begin{equation}
\frac{D \vec{u}}{D t} = - \frac{1}{ \rho_o  } \nabla_h p_b = \frac{ B_o R^3 z }{(x^2 + y^2 + z^2)^{5/2} } \big(x,y\big) \, .
\end{equation}
The environmental fluid above $z=0$ accelerates outward and is displaced away from the sphere while the fluid beneath $z=0$ accelerates inwards to fill the vacated space as depicted in Figure \ref{fig:sphere_plots}c.

\begin{figure*}
	\centering	\centerline{\includegraphics[width=7in]{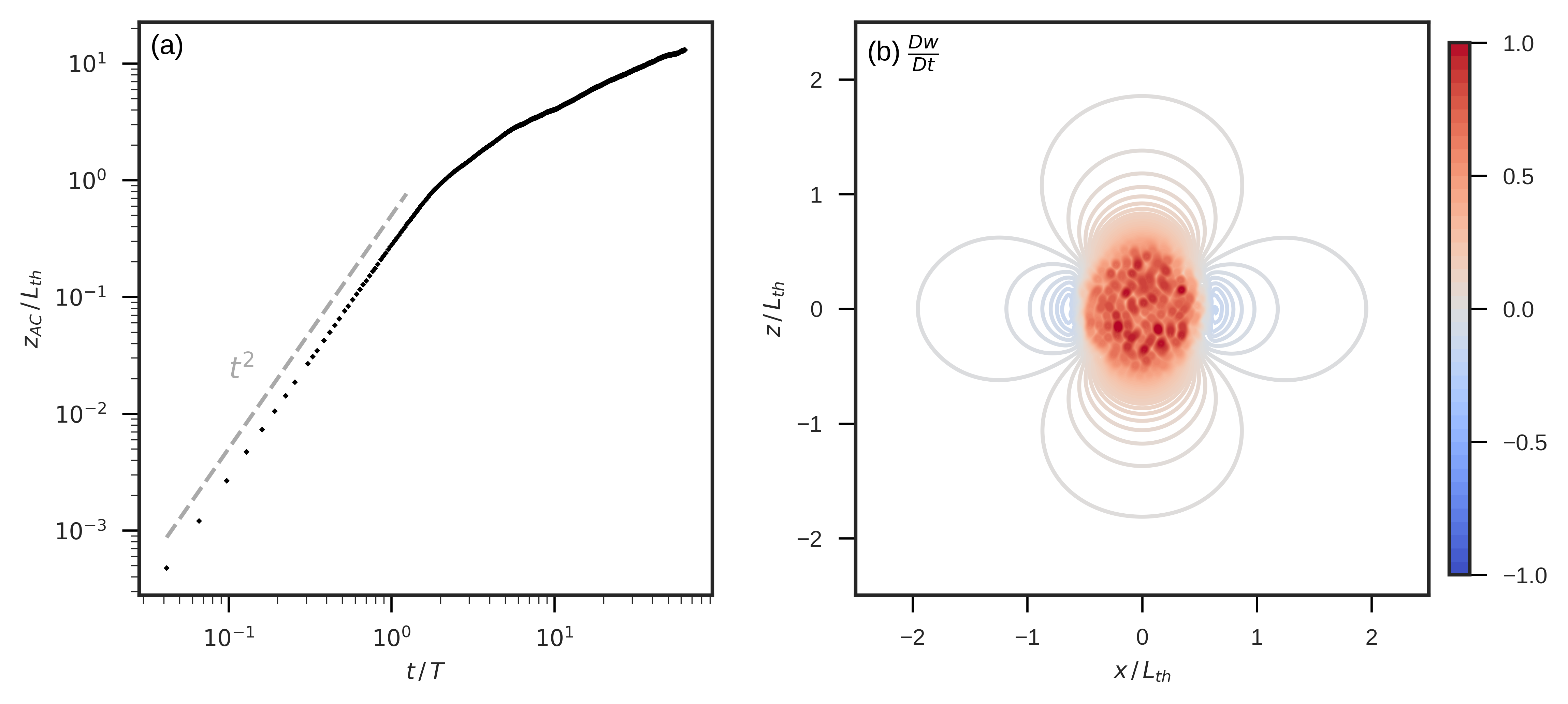}}
	\caption{a) log-log plot of the height of the anomaly versus time b) contour plot of the initial Lagrangian acceleration.}\label{fig:initial_regime_panel}
\end{figure*}

\section{Buoyant Motion From Rest}
\label{sec:motion_from_rest}
What is the relevance of the foregoing results? For a buoyancy anomaly accelerating from rest in a quiescent environment, the dynamic pressure is zero and the effective buoyancy gives the full Lagrangian acceleration. The motion will stay buoyancy-dominated until enough momentum has developed that the dynamic pressure force can no longer be neglected.

Since $L_{\text{th}}$ (the length of thermal) is the only length in the problem, it sets the length-scale of the buoyancy-dominated motion. The end of this regime occurs when $\nabla p_b \sim \nabla p_d$. Noting that $\nabla p_{\text{d}} \sim  \rho  W^2/L_{\text{th}} $ and $\nabla p_\text{b} \sim  \rho B_o$ due to the Poisson's equations, we have the following scalings for the buoyancy-dominated regime's duration and final velocity:    
\begin{align}
\tau  & \sim \sqrt{\frac{L_{\text{th}}}{B_o}} \, , \\
W & \sim  \sqrt{B_o L_{\text{th}}} \, .  \label{eqn:L_scaling}
\end{align}
During this buoyancy-dominated regime, a spherical thermal has near uniform effective buoyancy and thus achieves solid body motion with $\Delta z \approx \frac{1}{2} \beta t^2$. 

For a representative thermal with $L_{\text{th}} \sim .5$ km (\cite{romps2015sticky, hernandez2016numerical}), this parabolic motion would describe a thermal's rise through a typical boundary layer. With buoyancy in the characteristic range of $B=.01 - .1\  \mathrm{m/s^2}$, the scaling yields  $W_{\text{b}} \sim 2-7\ \mathrm{m/s}$ in close agreement with velocities observed for thermals in numerical studies \citep{romps2015sticky, hernandez2016numerical}. The above scalings are also in consonance with the idealized model for a spherical thermal put forward by \cite{escudier_maxworthy_1973}, who under several assumptions (e.g., that the thermal's motion is detrainement-free, drag from dynamic pressure is negligible, and the entrainment of ambient fluid is proportional to surface area and velocity) obtained asymptotic solutions for the thermal's trajectory. 

To validate this account of the buoyancy-dominated motion, we analyze the direct numerical simulations of a spherical thermal conducted in \cite{lecoanet_2018} using Dedalus,\footnote{For more information and links to the source code, see \url{dedalus-project.org}.} an open-source pseudo-spectral framework \citep{dedalus}. As detailed in Appendix A, we solve the non-dimensional Boussinesq equations for the turbulent motion of a heterogenous density anomaly rising through a neutrally-stratified environment. 

As shown in Figure \ref{fig:initial_regime_panel}b, the spatial structure of the instantaneous Lagrangian acceleration from rest matches the analytical $\beta$ result of Eqn. \ref{eqn:effective_buoyancy_sphere} (displayed in Figure \ref{fig:sphere_plots}). Let $\left<  \right>$ denote an average over the thermal's initial spherical volume. We find that $\left< \beta \right>/\left< B \right> = .65$, in close agreement with the theoretical prediction of 2/3. 

To verify the length-scale of the buoyancy-dominated regime, we track the center of the density anomaly through time by computing
\begin{equation}
\vec{x}_{\text{AC}}(t) = \frac{ \displaystyle\int_{\Omega}  \rho'(t) \, \vec{x} \, \, dV }{ \displaystyle\int_{\Omega} \rho'(t) \, \, dV } \, ,
\end{equation}
where $\Omega$ the full domain. As shown in Figure \ref{fig:initial_regime_panel}a, the buoyancy-dominated regime, characterized by $\Delta z \sim \beta t^2$ ascent, persists until the thermal rises a distance comparable to the thermal's own length, validating the analytical scaling of Eqn. \ref{eqn:L_scaling}.

\section{Effective Buoyancy of a Mature Thermal}
\label{sec:mature_thermal}
After the initial motion chronicled in the prior section, the thermal transitions to a turbulent regime distinguished by a \say{vortex ring} circulation. This structure, which loosely resembles the laminar Hill's vortex, consists of an ellipsoidal thermal that encases a rotating ring of more anomalously density; fluid ascends in the ring's interior and descends on the periphery \citep{scorer_1957, woodward_1959}. 

\begin{figure*}
	\centering
	\centerline{\includegraphics[width=6in]{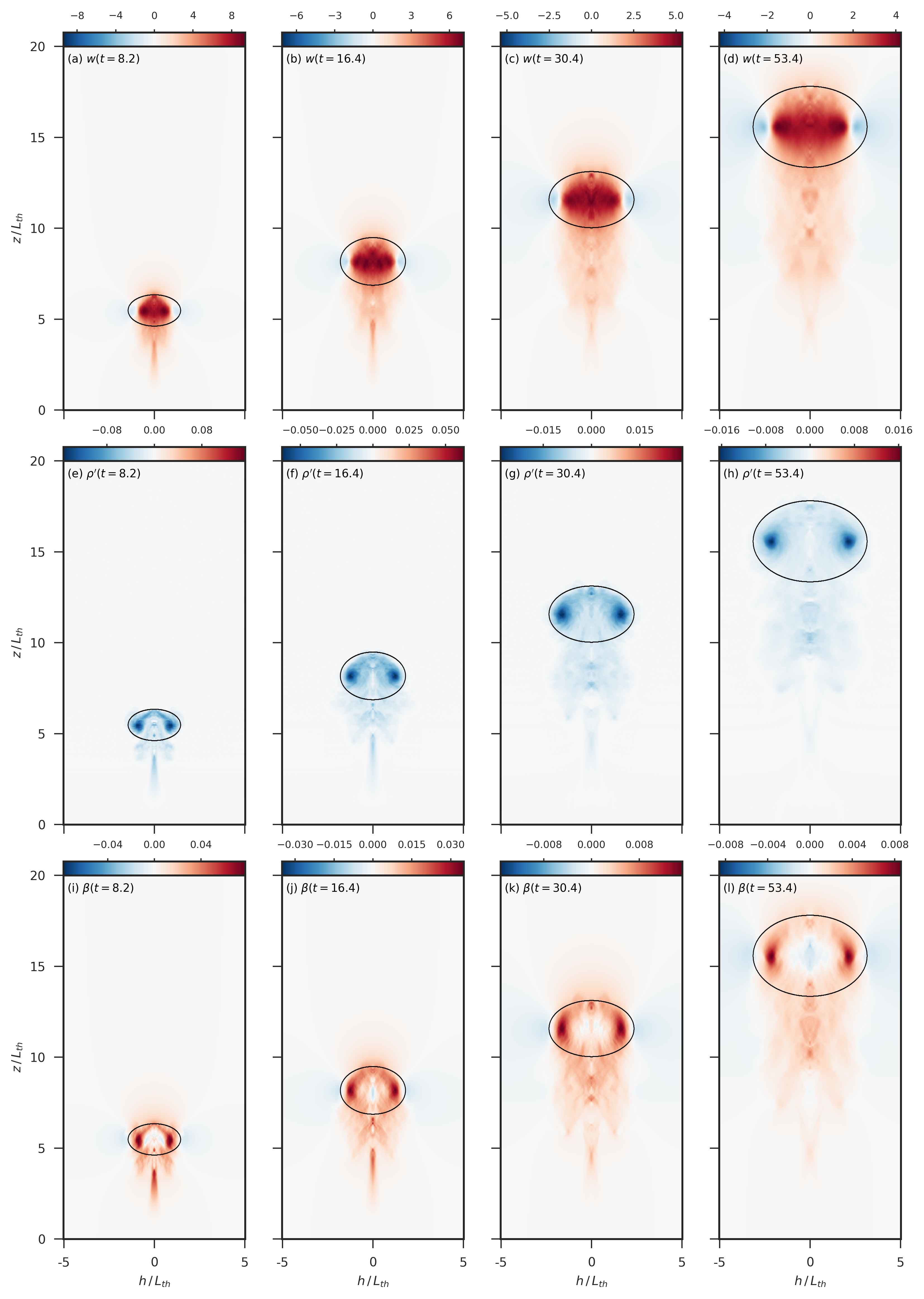}}
	\caption{Evolution of the azimuthally-averaged $w$, $\rho'$, and $\beta$ fields. Black contours show the identified boundaries of the actively convecting ellipsoidal thermals.}\label{fig:ellipsoid_panel}
\end{figure*}

Figure \ref{fig:ellipsoid_panel} demonstrates azimuthally-averaged $w$, $\rho'$, and $\beta$ fields for the ellipsoidal mature thermal. The $\beta$ distribution is found using the Poisson solver detailed in \cite{Jeevanjee2015}.  We observe that $\beta$ also is non-uniform in the thermal's interior: the concentration of buoyant fluid in the ring imposes a secondary structure with a pattern akin to the superposition of two buoyant spheres. Accordingly, $\beta$ is enhanced over the ring and opposite-signed in the center. We find that the presence of this ring of relatively undiluted fluid depends on the level of turbulence in our simulations. More laminar simulations in \citet{lecoanet_2018} find a much stronger ring. Considering that thermals in the atmosphere are far more turbulent than our simulations, this suggest that the $\beta$ field is closer to uniform in real thermals. 

Motivated by the findings of the prior section, we examine to what extent the mean $\beta$ of a mature thermal may be explained by considering a uniformly buoyant ellipsoid. With that goal in mind, we first derive the effective buoyancy of an ellipsoid via exploiting a mathematical equivalence between effective buoyancy and the equations of magnetostatics. 

\subsection{Magnetostatics Correspondence}
\label{subsec:magnetostatics}
Understanding the connection between magnetostatics and effective buoyancy requires familiarity with the concept of magnetization, which we outline below. Readers are referred to \cite{griffiths} or \cite{jackson} for a thorough treatment of the topic. 

Consider a small circular loop of wire through which a steady current flows. If the loop has vector area $\vec{a}$ and carries current $I$, taking  $\vec{a} \rightarrow 0$ while fixing the \emph{dipole moment} $\vec{m} = \vec{a}I$ yields the \emph{dipole magnetic field}. This miniature loop circuit is a conceptual model for the magnetic dipole moments generated by atomic electrons. A physical magnet is composed of many such dipole moments aligned in a common direction and described by the magnetic dipole moment per unit volume, known as the \textit{magnetization}, $\vec{M} = \frac{d \vec{m}}{dV}$. Inside a magnet that is uniformly magnetized in the vertical, we would have $\vec{M}  = M_o e_z$. 

We will consider the case of a permanently magnetized object, where the interior dipoles are aligned even in the absence of external magnetic fields. The total magnetic field $\vec{B}$ due to the magnet's permanent dipoles is found by first accomplishing the mathematically simpler task of determining the auxiliary $\vec{H}$-field, defined as 
\begin{equation}
\vec{H} = \frac{\vec{B}}{\mu_0} - \vec{M} \, . 
\end{equation}
In the absence of any free currents (i.e., currents following through wires), the curl and divergence of the above definition give
\begin{align}
\nabla \times \vec{H} &= 0  \\
\nabla \cdot \vec{H} &= - \nabla \cdot \vec{M} \, . 
\end{align} 
 Since $\vec{H}$ is curl-free, Helmholtz's theorem tells us that $\vec{H} = - \nabla \Phi_M$, where $\Phi_M$ is a scalar function known as the magnetic potential. We then have the Poisson's equation, 
\begin{equation}
 \nabla^2 \Phi_M  =  \nabla \cdot \vec{M} \, .\label{magnetic_poisson} 
\end{equation}
We regard $\nabla \cdot \vec{M} $ as the \say{ bound magnetic-charge}, which is non-zero only on the surface of the magnetized object. If the magnet is magnetized only in the vertical, then
\begin{equation}
\nabla^2 \Phi_M  =  \pd{M}{z} \,. 
\end{equation}
This equation is mathematically equivalent to  Eqn. \ref{eqn:poisson_buoyancy}, the Poisson's equation for the buoyancy perturbation pressure. In addition, $\Phi_M$ and $p_b$ share identical boundary conditions: both must be finite at the origin, vanish at $\infty$, and continuous over the boundary of a magnetic/fluid body.   

This mathematical equivalence spells out the following correspondence,  
\begin{equation}
\begin{split}
\text{Fluid}\quad \quad  & \text{Magnetostatics} \\
B\ \longleftrightarrow & \ M \\
\frac{ p_b }{\bar{\rho}} \ \longleftrightarrow &\ \Phi_M \\
- \frac{ \nabla p_b }{\bar{\rho}}\ \longleftrightarrow & \ \vec{H} \\
\vec{\beta} \ \longleftrightarrow & \ \frac{ \vec{B}}{\mu_0}
\end{split}
\label{magneto_corresp}
\end{equation}
The effective buoyancy of an arbitrarily-shaped fluid anomaly is mathematically equivalent to the magnetic field generated by a vertically magnetized body of the same shape. 

The classical theory of magnetized bodies is replete with analytical results. Poisson is attributed with finding the magnetic field inside a uniformly magnetized sphere and Maxwell gives the corresponding result for an ellipsoid in Article 437 of \cite{maxwell}. A modern treatment of the magnetic field inside a magnetized ellipsoid can be found in \cite{stoner1945xcvii}, and \cite{tejedor1995external} derives formulas for the external magnetic field presented in the following section. 

The magnetostatics correspondence also clarifies the relationship between effective buoyancy and induced mass. 
We have shown that a spherical buoyancy anomaly \emph{accelerates uniformly}, moving as if it were a solid body. This explains why its induced mass is equal to that of a solid sphere, and why the spatial structure of $\beta$ in Eqn. (\ref{eqn:effective_buoyancy_sphere}) is identical to that of the potential flow of a solid sphere translating at a steady velocity $U\vec{e}_z$ in an infinite fluid \citep[simply let the buoyant sphere evolve for time $\Delta t$ and set $U= (2 B_o/3) \Delta t$]{batchelor2000introduction}. For an ellipsoid, the effective buoyancy result presented in the next section also agrees with the induced mass formulas computed in Articles 114 and 373 of \cite{lamb}. Does an arbitrarily-shaped density anomaly also have a uniform effective buoyancy,and thus exactly mimic its solid body counterparts? 

Again, we turn to the magnetostatics literature. The effective buoyancy is uniform only if $\nabla p_b$ is uniform in the fluid anomaly's interior. In Article 437 of \cite{maxwell}. Maxwell proved that $\vec{H}$ -- the magnetostatics quantity analogous to $\nabla p_b$ -- is uniform in a magnetized body's interior only if the body's surface is quadratic. The only closed finite quadratic surface being the ellipsoid. Therefore, we conclude that only \emph{ellipsoidal} bodies have uniform $\beta$, and have induced mass coefficients equal to their solid body counterparts. 

\subsection{Ellipsoidal Thermal}
\label{subsec:ellipsoidal}

\label{subsubsec:spheriod_analytical_results}
Consider an azimuthally-symmetric ellipsoidal density anomaly of uniform Archimedean buoyancy $B_o$ and specified by major axes $(W,W,H)$. Let $\alpha = W/H$ be the aspect ratio of the ellipsoid with $\alpha < 1$ indicating a prolate ellipsoid and $\alpha > 1$ an oblate ellipsoid. Adapting the corresponding result for the magnetized ellipsoid presented in Article 437 of \cite{maxwell} and fully derived in \cite{stoner1945xcvii}, the buoyancy perturbation pressure in the interior of the ellipsoid takes the form,  
\begin{equation}
 p_b = f(\alpha) \rho_o B_o z \, , 
\end{equation} 
where we have defined 
\begin{equation}
 f(\alpha) = 
\begin{cases} 
\frac{\alpha^2}{1-\alpha^2} \left[ \frac{ \log \left( \frac{1+\sqrt{1-\alpha^2} }{\alpha} \right) }{\sqrt{1-\alpha^2}}  - 1\right] &  \alpha < 1 \\
\frac{1}{3}, &  \alpha = 1 \\
\frac{\alpha^2}{1-\alpha^2} \left[1 - \frac{1}{\sqrt{\alpha^2 -1}} \sin^{-1} \left( \frac{\sqrt{\alpha^2 -1}}{\alpha} \right) \right]  & \alpha > 1  \\
\end{cases} 
\label{eqn:f_alpha}
\end{equation}

Note that the oblate formula ($\alpha >1$) can be obtained by taking the real part of the prolate formula ($\alpha < 1$) and applying the identity $\sin^{-1}{x} = -i \ln(\sqrt{1-x^2} + ix)$. For reference, $f(\alpha)$ corresponds to the demagnetization factor $D(\mu)$ of Eqn. 4.2 in \cite{stoner1945xcvii} with $\alpha = \mu$.

\begin{figure}
	\centering	\centerline{\includegraphics[width=3in]{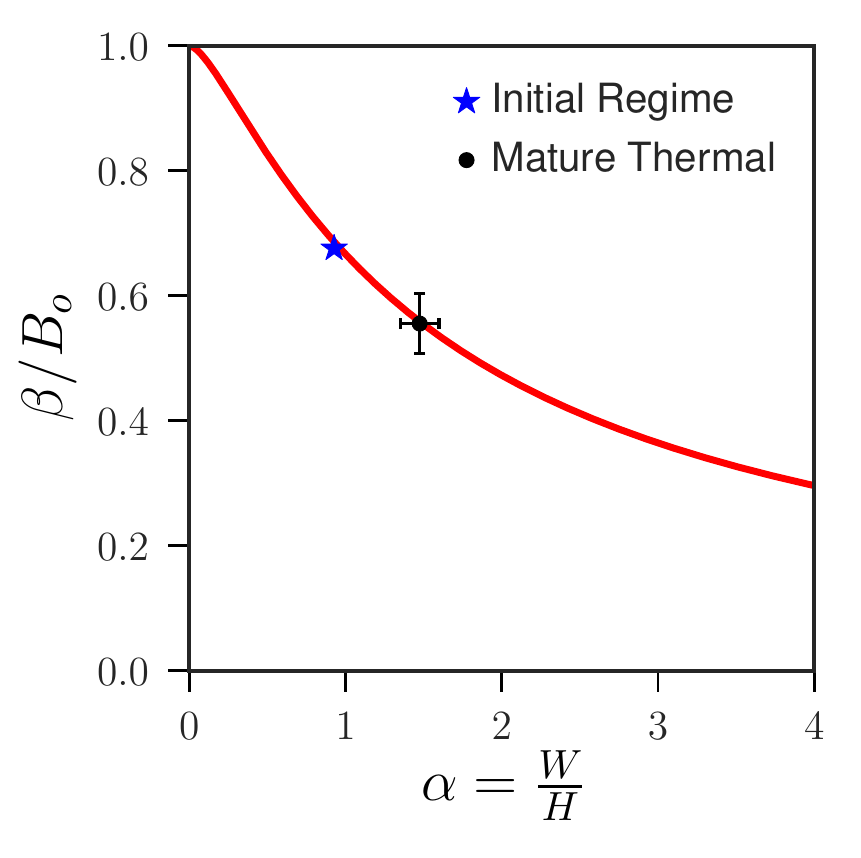}}
	\caption{Comparison between $\beta$ averaged over the ellipsoidal thermal as a function of the diagnosed aspect ratio and the analytical formula (Eqn. \ref{eqn:beta_spheroid}).  Thermal identification was conducted every .1 $T$, yielding 5 snapshots of the initial regime and 70 snapshots of the mature thermal. Bars indicate the standard deviation of the data for each regime (error bars for the initial regime are smaller than marker). } \label{fig:beta_vs_B_comparison}
\end{figure}

As in the sub-case of the sphere, $p_b$ is a linear function of $z$ inside the anomaly, and thus the whole ellipsoid uniformly accelerates in the vertical with an effective buoyancy of 
\begin{equation}
 \beta_{\text{ellipsoid}}(\alpha) = (1 - f(\alpha)) B_o \, , \label{eqn:beta_spheroid}  
\end{equation}
as shown in Figure \ref{fig:beta_vs_B_comparison}. This formula is the quantitative foundation for the observation of \cite{houze2014cloud} that wide pancake-shaped thermals are severely inhibited by their environment with $\beta \approx 0$, while narrow pencil-shaped anomalies impart negligible momentum to their environment yielding $\beta \approx B$.

Importantly, it also enables us to move beyond the limiting cases and understand the regime relevant for atmospheric thermals. \cite{hernandez2016numerical} found evidence for slightly prolate thermal, while \cite{scorer_1957, woodward_1959, damiani_2006} observe moderately oblate thermals consistent with our simulations, which have $\alpha \approx 1.4$. 

\subsection{Comparison to Simulation} 
\label{subsec:comparison}
Equipped with the idealized solution (Eqn. \ref{eqn:beta_spheroid}), we investigate its potential relevance to the simulated ellipsoidal thermals. Rigorous comparison to the simulation results requires introducing an objective method to identify the thermal's boundary. To achieve this, we implement a simple $\rho'$-based approach as described in Appendix B. Example identification results are shown in Figure \ref{fig:ellipsoid_panel}.  

Thermal identification enables us to compute the volume average of $\beta$ over the thermal, again denoted as $\left< \beta \right> $. Figure \ref{fig:beta_vs_B_comparison} compares $\left< \beta \right>/\left< B \right>$ to the analytical result of Eqn. \ref{eqn:beta_spheroid}, evaluated with the thermal's aspect ratio. The analytical curve is found to accurately explain the effective buoyancy over the mature thermal's ascent. In particular, the time-mean $\left<\beta \right>$ of the turbulent thermal is within $.01 B_o$ of the analytical result --- a remarkable agreement. This finding allows us to go beyond scaling arguments and make a first analytical step towards rigorously understanding the forces acting on a turbulent thermal. 

\begin{figure}
	\centering	\centerline{\includegraphics[width=3in]{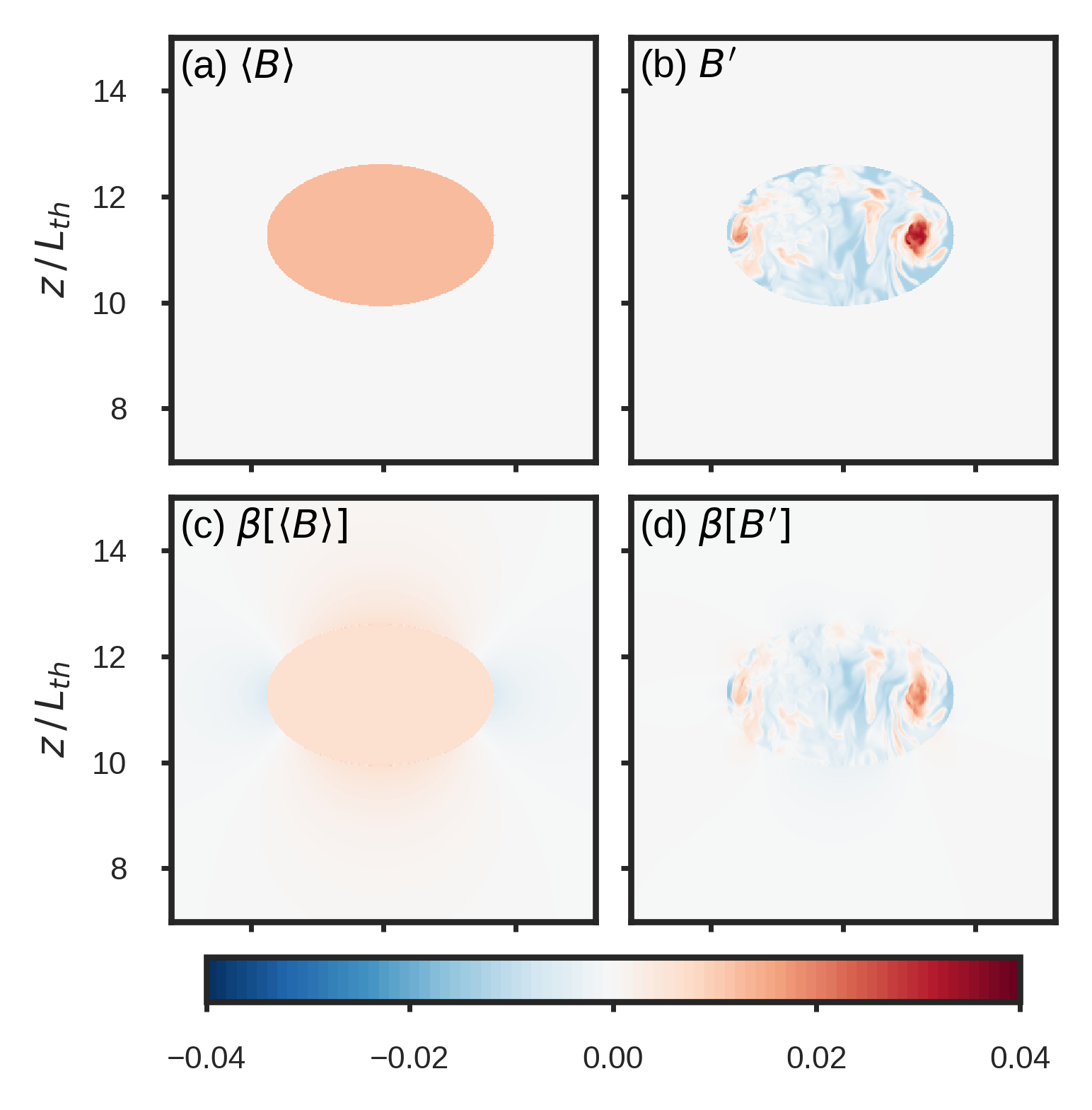}}
	\caption{a) vertical cross section at $t = 28.7$  of $\left< B  \right>$, the volume average of the thermal's Archimedean buoyancy, b) $B' = B - \left< B  \right>$, the departures from the mean ellipsoidal buoyancy, c) effective buoyancy generated by $\left< B  \right>$ distribution, d) effective buoyancy generated by $B'$ distribution. See text for discussion.  }\label{fig:mean_vs_prime}
\end{figure}

A lingering question, however, is: why do we find such close agreement despite the heterogeneous ring structure present in the $B$ field? Consider decomposing the thermal's Archimedean buoyancy field according to  $B = \left<B\right> + B'$. By definition, we have that $\left<B'\right>=0$, and thus $B'$ contains opposite-signed regions that cancel when averaged over the thermal. Taking advantage of the linearity of Poisson's equation, $\beta$ admits the decomposition $\beta = \beta\left[\left<B\right>\right] + \beta\left[B'\right]$, where $\beta\left[\left<B\right>\right]$ is the effective buoyancy generated by a uniform ellipsoid of constant Archimedean buoyancy $\left<B\right>$. Likewise, $\beta[B']$ is the effective buoyancy resulting from the heterogeneous $B'$ field. See Figure \ref{fig:mean_vs_prime} for a plot of these quantities. As expected, $\beta\left[\left<B\right>\right]$ is constant in the ellipsoid and described by the analytical result of the prior section. 

We observe that the $\beta\left[B'\right]$ field inherits the heterogeneous features of the $B'$ field and preserves the distribution of opposite-signed regions. As a consequence, if we average $\beta\left[B'\right]$ over the thermal, cancellation between the opposite-signed regions yields a value three orders of magnitude smaller than $\beta\left[\left<B\right>\right]$. Therefore, we find that the departures from $\left<B\right>$ make a negligible contribution to $\left< \beta \right>$, supplying the agreement observed in Figure \ref{fig:beta_vs_B_comparison}. 

\section{Discussion}
\label{sec:discussion}

Our main findings are as follows:

\begin{itemize}
	\item There exists an exact correspondence between magnetostatics and the fluid dynamics of effective buoyancy (Eqn. \ref{magneto_corresp})
        \item This correspondence supplies analytical formula for the effective buoyancy of spherical and ellipsoidal anomalies---the only fluid geometries that experience a uniform buoyant acceleration, and thus have virtual mass coefficients identical to their solid-body counterparts. 
        \item The effective buoyancy of a heterogenous, turbulent thermal is captured by an analytical function of the thermal's aspect ratio (Fig. \ref{fig:beta_vs_B_comparison}). 
\end{itemize}

The above result enables the computation of a turbulent thermal's buoyant acceleration from the  coarse  geometry of the density field, and thus can be applied to lower resolution observational or model data.  Another application of these results is to cumulus parameterizations, and specifically those that parameterize the vertical velocity equation. As summarized by \cite{de2012parameterization}, a wide range of virtual mass coefficients are employed in the literature. However, the results presented here, and in particular Fig. \ref{fig:beta_vs_B_comparison}, suggest that the virtual mass coefficients much different from 2/3 are not justified by thermal-based reasoning. 

A limitation of this study is that our simulations do not account for water vapor or a stratified environment. This poses a problem, however, only if a \say{wet} thermal in a stratified fluid deviates from the coarse ellipsoidal geometry necessary to apply our analysis. Fortunately, atmospheric observations (\cite{damiani_2006}) and \say{wet} stratified simulations (\cite{sherwood_2013, romps2015sticky, hernandez2016numerical}) find thermals that possess the necessary ellipsoidal shape.

Note that our analysis emphasized the buoyancy perturbation pressure, but the dynamic pressure force is equally important to the circulation of an ellipsoidal vortex ring. While there has been recent progress in visualizing and parameterizing the dynamic pressure force \citep{peters_2016}, a deeper quantitative understanding is still lacking. In particular, we would benefit from  analytical solutions to \eqref{eqn:poisson} that reveal the dynamic pressure's spatial pattern and mean effect on a mature thermal. Future research should aim to address these questions. 

\acknowledgments
NT thanks Kirk McDonald for discussion of the magnetostatics content, and for bringing Maxwell's proof to his attention. The authors thank three anonymous reviewers in addition to Howard Stone, Aaron Match, and David Romps for helpful discussions and feedback. NJ is supported by a Harry Hess fellowship from the Princeton Geosciences Department. DL is supported by a PCTS fellowship and a Lyman Spitzer Jr.~fellowship. Computations were conducted with support by the NASA High End Computing (HEC) Program through the NASA Advanced Supercomputing (NAS) Division at Ames Research Center on Pleiades with allocations GID s1647 and s1439.
\appendix[A]
\appendixtitle{Simulation}
\label{sec:simulation}

We briefly outline the details of the simulations and refer readers to \cite[Section 2]{lecoanet_2018} for a complete treatment. We non-dimensionalize according to $\hat{\vec{x}} = \vec{x}/L_{\text{th}} $, $\hat{\vec{v}} = \vec{v}/V_{\text{th}} $, and $\hat{\rho'} = \rho'/ \alpha \Delta T$, where the thermal has characteristic length $L_{\text{th}}$, velocity $V_{\text{th}}$, and  anomalous density $\alpha \Delta T$ (here, $\alpha$ is the coefficient of thermal expansion). Non-dimensionalizing the Boussinesq equations provides the definition of the thermal's characteristic velocity,
\begin{equation}
V_{\text{th}} = \sqrt{ \frac{ g L_{\text{th}} \alpha \Delta T}{\rho_o } } \, . 
\end{equation}
The solution to the non-dimensional equations is then completely described by the Reynold's number, 
\begin{equation}
\text{Re} = \frac{V_\text{th} L_\text{th}}{\nu} \,,
\end{equation} 
and the Prandtl number,
\begin{equation}
\text{Pr} = \frac{\kappa}{\nu} \, , 
\end{equation}
where $\kappa$ is the thermal diffusivity. We take $\text{Pr}=1 $ and $\text{Re} = 2/\sqrt{10} \times 10^4 \approx 6300$.

To resolve the small scale features of the initial acceleration field shown in Fig. \ref{fig:initial_regime_panel}b, we initialize the sphere in the center of a cubic domain of length $5 L_\text{th}$, and represent the solution with $(1024)^3$ Fourier modes. To simulate the ascent of a mature thermal, we initialize the sphere above the bottom of a rectangular domain of height 20$L_\text{th}$ and horizontal extent $10L_\text{th}$, and use 512 modes in the horizontal directions and 1024 modes in the vertical.

The spherical anomaly is seeded with a random heterogenous $\hat{\rho'}$ distribution such that the average of $\hat{\rho'}$ over the sphere is $-1$. In the smaller domain (larger domain) simulation, the root-mean-square of $\hat{\rho'}$ is $.48$ ($.21$). Both simulations are run until the anomaly collides with the top of the domain. In the larger domain simulation, analysis was conducted on the thermal's ascent (shown in Figure \ref{fig:ellipsoid_panel}) from $z = 6 L_{\text{th}}$ to $z = 16 L_{\text{th}}$, during which the thermal exhibits ellipsoidal vortex motion and does not interact with the domain top.

\appendix[B]
\appendixtitle{Thermal Tracking}
\label{sec:thermal_tracking}

\begin{figure}
	\centering
	\centerline{\includegraphics[width=3in]{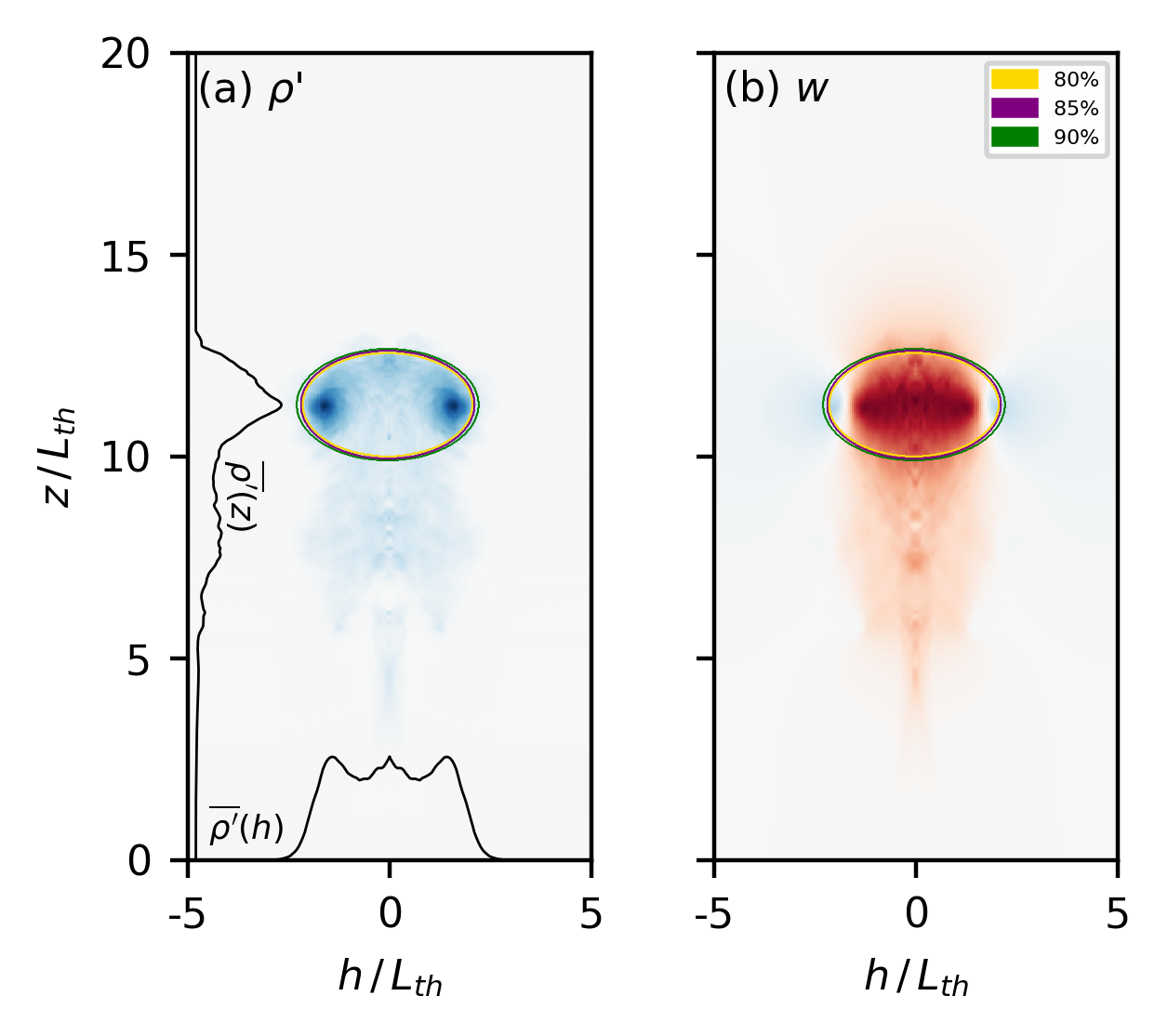}}
	\caption{The azimuthally-averaged a) $\rho'$ and b) $w$ field at $t = 28.7$ with identified ellipsoidal boundaries corresponding to $80\%$, $85\%$, and $90\%$ $|\rho'|$ reduction thresholds. The column-averaged $\bar{\rho'}(h)$ and row-averaged $\bar{\rho'}(z)$ are also shown.}\label{fig:azi_detection}
\end{figure}

To minimize complexity, we solve for the ellipsoidal thermal volume by identifying the major axes of an azimuthally-symmetric ellipsoid via a fixed threshold on $\rho'(h,z)$, the azimuthally-averaged density anomaly field, where $h=\sqrt{x^2 +y^2}$ is the horizontal radial coordinate.  

We first locate the central vertical axis of the thermal by performing a vertical velocity-weighted average of $(x,y)$ on each vertical level. After computing the azimuthal average of $\rho'(h,z)$ about this central axis, we reduce the dimensionality again by vertically averaging $\rho'(h,z)$ over each column to yield $\bar{\rho'}(h)$, and separately averaging $\rho'(h,z)$ across each row to give $\bar{\rho'}(z)$. Both averages are shown in Figure \ref{fig:azi_detection}.

The extent of the ellipsoid's vertical axis is set by the height above the ellipsoid's center where $\bar{\rho'}(z)$ is reduced by $85\%$ from its maximum value. Likewise, the horizontal axis is set by the radial distance at which $\bar{\rho'}(h)$ is reduced by $85\%$ from its maximum value.  To minimize subjectivity, we conduct a sensitivity analysis and identify the ellipsoid with alternate choices of an $80 \%$ and $90 \%$ reduction threshold. Figure \ref{fig:azi_detection} reveals that identification results are robust to variations of the $\rho'$ threshold. More sophisticated thermal tracking methods exist (e.g., \cite{romps2015sticky}), but we believe this rudimentary algorithm does a sufficient job of detecting the prominent ellipsoidal structure of the thermal.

 \bibliographystyle{ametsoc2014}
 \bibliography{buoyancy_revision_references}

\end{document}